# New exact solutions of the Dirac equation of a charged particle interacting with an electromagnetic plane wave in a medium

Sándor Varró

Wigner Research Centre for Physics, Hungarian Academy of Sciences,
Institute for Solid State Physics and Optics, Budapest, Hungary
E-mail: varro.sandor@wigner.mta.hu

**Abstract.** Exact solutions are presented of the Dirac equation of a charged particle moving in a classical monochromatic electromagnetic plane wave in a medium of index of refraction $n_m < 1$. The found solutions are expressed in terms of new complex trigonometric polynomials, which form a doubly infinite set labeled by two integer quantum numbers. These quantum numbers represent quantized spectra of the energy-momentum components of the charged particle along the polarization vector and along the propagation direction of the applied electromagnetic plane wave field (which is considered as a laser field of arbitrary high intensity propagating in an underdense plasma). The found solutions may serve as a basis for the description of possible quantum features of mechanisms of acceleration of electrons by high-intensity laser fields.

## 1. Introduction

The Volkov states have long been an important tool in the theoretical description of fundamental processes taking place in strong laser fields (Fedorov 1997). They, and their scalar or non-relativistic variants are exact solutions of the Dirac equation (Volkov 1935) and Klein-Gordon equation (Gordon 1927) or the Schrödinger equation (Keldish 1964, Bunkin and Fedorov 1965) of an electron or other charged particle moving in a true plane electromagnetic wave in vacuum, or in dipole approximation, respectively. On the basis of these wave functions, the interaction with the laser field in the initial, final or/and in the immediate states of an electron can be taken into account exactly, i.e. up to 'infinite order' (in the terminology of perturbation theory). Recently there has been a renoved interest in the physical applications and mathematical properties of these states, concerning the theory of laser-matter interactions. In the relativistic regime the closed analytic form of these set of states rely on the vacuum dispersion relation of the classical electromagnetic (EM) plane wave, which represents the laser field. Many mathematical details (e.g. orthogonality and completeness) of these solutions have been investigated during decades (see Brown and Kibble 1964, Eberly 1969, Neville and Rohrlich 1971, Ritus and Nikishov 1979, Bergou and Varró 1980), and even nowadays their study is undergoing a certain 'renaissance' (see e.g. Zakowicz 2005, Salamin *et al* 2006, Musakhayan 2008, Ehlotzky *et al* 2009, Boca and Florescu 2009, 2010). We note that the solutions of the Dirac equation can also be expressed in a closed form for a quantized electromagnetic plane wave, as has first been shown by Bersons (1967).

Besides the generalizations of these quantized states (Fedorov and Kazakov 1973, Bersons and Valmanis 1973), they have also been used for treating the nonlinear Compton scattering beyond the semiclassical approximation (Bergou and Varró 1981*b*). The generalization of the Volkov states with a quantized radiation have also been used to describe multiphoton stimulated Bremsstrahlung in the nonrelativistic regime (Bergou and Varró 1981*a*), which process has been reinvestigated (Burenkov and Tikhonova 2010) from the point of view of possible nonclassical effects. The appearance of various versions of entanglement in strong-field laser-matter interactions has also been investigated by Fedorov *et al* (2006) for brakeup processes, and by Varró (2008, 2010*a-b*) for the photon-electron system itself. In the present paper we shall stay in the frame of the semiclassical description of the radiation field, and will not discuss many-particle correlations and entanglement in the presence of strong laser fields.

If the EM field propagates in a medium (e.g. with a real index of refraction $n_m < 1$ or $n_m > 1$), or the strong laser field is modelled by a time-periodic electric field, then the by now studied solutions of the relativistic wave equations can be expressed by the solutions of the corresponding Mathieu or Hill equations (see Narozhny and Nikishov 1974, Becker 1977, Cronström and Noga 1977). Such an analysis have for example been applied in the mathematical study of pair creation in strong fields (Nikishov and Ritus 1967, Nikishov1970). This phenomenon has received a renoved interest recently (see e.g. Narozhny *et al* 2004, Popov 2004, Dunne 2004, 2009), owing to the considerable technological development of extremely high power and ultrashort laser systems (see e.g. Krausz and Ivanov 2009, Mourou *et al* 2006 and 2013). Concerning the mathematical description, except for very special cases (see e.g. Feldberger and Marburger 1975), the solutions cannot be expressed in a closed finite form; in the simplest case they are related to the (trancendental) Mathieu functions.

In the present paper we show that there are exact closed form solutions of the Dirac equation of a charged particle moving in a monochromatic classical plane EM field in a medium of index of refraction $n_m < 1$. It is well-known that such a radiation field can be transformed to a homogeneous oscillating electric field (Narozhny and Nikishov 1974, Becker 1977), by going over to a suitably chosen Lorentz frame, thus our considerations are relevant for discussing this type of interaction, too. The solutions to be derived are proportional with finite complex trigonometric polynomials whose arguments are integer multiples of the EM wave's phase, and they form a doubly infinite discrete set of solutions labeled by two integer quantum numbers. This latter property is a completely new feature in comparison with both the Volkov states (in vacuum) parametrized by continouos four momenta, and with the Mathieu type solutions characterized by 'stability charts' resulting in a band structure of the allowed parameter.

In section 2 we construct the differential equations for the scalar coefficient functions of the constant bispinor basis, from which the complete Dirac bispinor is built up. In the course of the derivation we also make a brief comparison with the method leading to the usual Volkov states or to the Mathieu-type wave functions. Section 3 is devoted to the determination of the complex polynomial solutions, which are associated to the eigenvalue problem of special finite tri-diagonal matrices. The solutions are labeled by two integer quantum numbers, which represent a quantized spectrum of the electron's momentum components along the (linear) polarization direction and along the propagation of the laser field. In section 4 we shall show the main mathematical properties of the found solutions, and few physical implications associated to these solutions will be discussed. Besides, some numerical examples will also be presented, for the momentum spectrum, the temporal behaviour and harmonic structure of the wave functions, just for illustration purposes. In section 5 a brief summary with conclusions closes our paper.

## 2. Construction of the wave equations for the scalar coefficient functions of the Dirac bispinors

In an external electromagnetic field characterized by the four-vector potential $A(x)$ the Dirac equation of a spinor particle of charge $e$ and of mass $m$ has the form[1]

$$[\gamma \cdot \Pi - \kappa]\psi = 0, \quad \Pi \equiv i\partial - \varepsilon A, \quad \varepsilon = e/\hbar c, \quad \kappa = mc/\hbar, \quad (1)$$

where $c$ is the velocity of light in vacuum, and $\hbar$ is Planck's constant divided by $2\pi$. In (1) we have introduced the operator of the kinetic four-momentum $\Pi$ of the charged particle. In a medium of index of refraction $n_m$, a general transverse electromagnetic plane wave of wave vector $k$ can be represented by a vector potential

$$A(x) = e_1 A_1(\xi) + e_2 A_2(\xi), \quad \xi = k \cdot x, \quad k \cdot A = 0, \quad k = \{k^\mu\} = k^0(1, n_m e_3), \quad \{e_{1,2}^\mu\} = (0, e_{1,2}), \quad (2)$$

where $\{e_1, e_2, e_3\}$ form a right system of mutually orthogonal unit vectors. In principle, the scalar functions $A_{1,2}(\xi)$ in (2) may have arbitrary form (satisfying, of course, certain regularity conditions). For

$$\boldsymbol{\alpha} = \begin{pmatrix} 0 & \boldsymbol{\sigma} \\ \boldsymbol{\sigma} & 0 \end{pmatrix}, \quad \beta = \begin{pmatrix} 1 & 0 \\ 0 & -1 \end{pmatrix}, \quad \boldsymbol{\gamma} = \begin{pmatrix} 0 & \boldsymbol{\sigma} \\ -\boldsymbol{\sigma} & 0 \end{pmatrix}, \quad \sigma_x = \begin{pmatrix} 0 & 1 \\ 1 & 0 \end{pmatrix}, \quad \sigma_y = \begin{pmatrix} 0 & -i \\ i & 0 \end{pmatrix}, \quad \sigma_z = \begin{pmatrix} 1 & 0 \\ 0 & -1 \end{pmatrix}.$$

---

[1] The Minkowski metric tensor $g_{\mu\nu} = g^{\mu\nu}$ has the components $g_{00} = -g_{ii} = 1$ ($i = 1, 2, 3$) and $g_{\mu\nu} = 0$ if $\mu \neq \nu$ ($\mu, \nu = 0, 1, 2, 3$). The scalar product of two four-vectors $a$ and $b$ is $a \cdot b = g_{\mu\nu}a^\mu b^\nu$, i.e. $a \cdot b = a_\nu b^\nu = a^0 b^0 - \mathbf{a} \cdot \mathbf{b}$, where $\mathbf{a} \cdot \mathbf{b}$ is the usual scalar product of three-vectors $\mathbf{a}$ and $\mathbf{b}$. Space-time coordinates are denoted by $x^\mu$, where $x = \{x^\mu\} = (ct, \mathbf{r})$. The four-gradient is $\partial = \{\partial^\mu\} = (\partial/\partial ct, -\partial/\partial \mathbf{r})$, and $\partial_\mu = \partial/\partial x^\mu$. In the standard representation the Dirac matrices $\boldsymbol{\alpha} = (\alpha_x, \alpha_y, \alpha_z)$ and $\beta$ have the form

In the first three equations the „0" and „1" denote 2×2 zero and unit matrices, respectively. In the last three equations $\sigma_{x,y,z}$ are the usual 2×2 Pauli matrices. The $\gamma$ matrices are defined as $\gamma^{1,2,3} = \gamma_{x,y,z} = \beta\alpha_{x,y,z}$ and $\gamma^0 = \beta$, their commutation relations are $\gamma^\mu \gamma^\nu + \gamma^\nu \gamma^\mu = 2g^{\mu\nu}$, and $\gamma^0 \gamma_\mu^+ \gamma^0 = \gamma_\mu$ (cf Bjorken and Drell (1964)), where $\gamma_\mu^+$ denotes the adjoint (transposed conjugate) of $\gamma_\mu$ (cf Bjorken and Drell 1964).

a purely monochromatic field $A_{1,2}(\xi)$ are simple sine and cosine functions and $k^0 = \omega_0/c$, where $\omega_0 = 2\pi\nu_0$ is the circular frequency. As has long been shown (Volkov 1935), in the case when $n_m = 1$, i.e. $k^2 = 0$, the solutions of the above Dirac equation can be expressed in a simple closed form; these are the Volkov states. Various physical applications and mathematical details of these states (concerning the boundary conditions, positive and negative energy solutions, orthogonality and completeness) can be found for instance in Brown and Kibble (1964), Eberly (1969), Bergou and Varró (1980), Zakowicz (2005), Boca and Florescu (2010).

Towards the solution of the Dirac equation for the bispinor $\psi$, it is customary to go over to the second-order Dirac equation for a new bispinor $\Psi$, by using the Ansatz

$$\psi = (\gamma \cdot \Pi + \kappa)\Psi,$$

$$\left[\Pi^2 - \kappa^2 - \frac{1}{2}\varepsilon\sigma \cdot F\right]\Psi = 0, \quad \sigma \cdot F = \sigma_{\mu\nu}F^{\mu\nu}, \quad \sigma_{\mu\nu} = \frac{i}{2}[\gamma_\mu, \gamma_\nu], \quad F^{\mu\nu} = \partial^\mu A^\nu - \partial^\nu A^\mu, \quad (3)$$

where we have introduced the spin tensor $\sigma_{\mu\nu}$ and the electromagnetic field strength tensor $F^{\mu\nu}$. One can easily show that the matrix part $\sigma \cdot F$ on the left hand side of the second order Dirac equation has the explicit form $\sigma \cdot F = 2i(\gamma \cdot k)[\gamma \cdot A'(\xi)]$, where the prime denotes derivative with respect to $\xi$. By considering a modified plane wave solution, labelled by a four-momentum parameter $p_\mu$, we have

$$\Psi = \Psi_p(\xi)\exp(-ip \cdot x),$$

$$-k^2\frac{d^2\Psi_p}{d\xi^2} + 2ik \cdot p\frac{d\Psi_p}{d\xi} + \left[p^2 - \kappa^2 - 2\varepsilon p \cdot A + \varepsilon^2 A^2 - \frac{1}{2}\varepsilon\sigma \cdot F\right]\Psi_p = 0, \quad \xi = k \cdot x. \quad (4)$$

In vacuum ($n = 1$, $k = k_0(1, e_3)$, $k^2 = 0$) the factor of the second derivative in (4) is zero, and the matrix part is nilpotent, $(\sigma \cdot F)^2 = 0$, because $(\gamma \cdot k)^2 = k^2 = 0$ in this case. The remaining first order equation can be directly integrated, yielding the Volkov states. The mathematical complexity of the problem is largely increased if one considers the interaction with a plane wave propagating in a medium of index of refraction $n_m \neq 1$. This is due to the presence of the non-vanishing second derivative in (4), since now $k^2 = k_0^2(1 - n_m^2) \neq 0$. Anyway, in this case we can eliminate the first derivative in (4), or a priori use the following modified Ansatz at the outset,

$$\Psi = \Psi_p^{(\pm)}(\xi)\exp\left\{\mp i\left[p - \frac{(k \cdot p)}{k^2}k\right] \cdot x\right\} \equiv \Psi_p^{(\pm)}(\xi)\exp[\mp i(-\hat{p}\hat{x} - p_1 x_1 - p_2 x_2)], \quad k^2 > 0, \quad (5a)$$

$$\hat{p} \equiv \hat{k} \cdot p/(k^2)^{1/2}, \quad \hat{x} \equiv \hat{k} \cdot x/(k^2)^{1/2}, \quad k = \{k^\mu\} = k^0(1, n_m e_3), \quad \hat{k} = \{\hat{k}^\mu\} = -k^0(n_m, e_3), \quad (5b)$$

$$\frac{d^2\Psi_p^{(\pm)}}{d\xi^2} + \frac{1}{-k^2}\left[-p_\xi^2/k^2 + p^2 - \kappa^2 \mp 2\varepsilon p \cdot A + \varepsilon^2 A^2 - \frac{1}{2}\varepsilon\sigma \cdot F\right]\Psi_p^{(\pm)} = 0, \quad p_\xi \equiv (k \cdot p), \quad (5c)$$

where the ambient sign $\mp$ refers to the 'positive and negativ energy solutions'. The plane wave factor in the first equation in (5a) has also been written as $\exp[\mp i(-\hat{p} \cdot \hat{x} - p_1 x_1 - p_2 x_2)]$, where $p_1 = e_1 \cdot p$ and $p_2 = e_2 \cdot p$ are the transverse momentum components. We have introduced the 'complementary wave vector' $\hat{k} = -k^0(n_m, e_3)$ ($\hat{k}^2 = -k^2$ and $\hat{k} \cdot k = 0$), with the help of which one can derive

$$p = \frac{(k \cdot p)}{k^2} k - \frac{(\hat{k} \cdot p)}{k^2} \hat{k} - (p \cdot e_1)e_1 - (p \cdot e_2)e_2, \quad \left[ p - \frac{(k \cdot p)}{k^2} k \right] \cdot x = -\hat{p}\hat{x} - p_1 x_1 - p_2 x_2. \qquad (5d)$$

This kind of expansion can be performed for any other four-vectors, of course. If $n_m^2 < 1$, then $\hat{k}$ is space-like, and $\hat{p} \equiv \hat{k} \cdot p / (k^2)^{1/2}$ plays the role of a momentum component conjugated to the 'position variable' $\hat{x} \equiv \hat{k} \cdot x / (k^2)^{1/2}$ (see e.g. Becker 1977). Thus, the Ansatz in (5a) means a separation of variables, where $\hat{k} \cdot p$, $p_1$ and $p_2$ are 'three-momentum type parameters'. Accordingly, $p_\xi = (k \cdot p)$ may be said to be an 'energy type parameter'. One should keep in mind that $p_\xi$ does not explicitly show up in the solution, because $(p_\xi / k^2) k_\mu$ is subtracted from $p_\mu$, as is shown by the second equation of (5d).

At this point we would like to note that for a circularly polarized monochromatic plane wave one has $A_{cir}(x) = A_0(e_x \cos \xi + e_z \sin \xi)$ and $A_{cir}^2 = -A_0^2 = $ constant. By neglecting the spin interaction term $\sigma \cdot F$ in (5c) (or by *a priori* considering a Klein-Gordon particle of wave function $\Phi_p^{(\pm)} \exp(\mp ip \cdot x)$), we immediately obtain a *Mathieu equation* for the scalar modulation function $\Phi_p^{(\pm)}$ (see e.g. Becker 1977, Cronström and Noga 1977). In case of a linearly polarized monochromatic wave $A_{lin}(x) = A_0 e_x \cos \xi$ (with $A_{lin}^2 = -A_0^2 (1 + \cos 2\xi)/2 \neq $ constant), the scalar modulation function $\Phi_p^{(\pm)}(\xi)$ would satisfy the so-called *Whittaker–Hill equation* (or Hill's three-term equation, see Arscott 1964). If one would attempt to solve this equation in terms of a trigonometric series, that procedure would lead to five-term recurrence relations between the coefficients, in contrast to the three-term expressions encountered with the Mathieu equation. Thus, the standard procedure (the technique of continued fractions) cannot be taken over from the theory of Mathieu equation. In their study of pair production by a periodic electric field, Narozhny and Nikishov (1974) also considered the sub-case $p \cdot A = 0 = p_x$, when the resulting differential equation can also be brought to a Mathieu equation.

In the rest of the present paper we shall study the general equation (5c), however, we emphasize that we shall not carry out a complete analysis of this equation, which has a quite unexplored infinite set of transcendental solutions. We shall rather restrict our analysis to the special class of solutions, proportional with polynomial expressions (finite complex Fourier sums), which form a subset of all solutions. The exceptional feature of these solutions is that they form a doubly infinite countable set, corresponding to discrete values of the transverse and longitudinal momentum parameters.

Henceforth, in equation (5c) we specialize $A = A(\xi)$ to represent a monochromatic linearly polarized plane wave,

$$A(x) = e_x A_0 \cos\xi, \quad \xi = k \cdot x, \quad \{k^\mu\} = \frac{\omega_0}{c}(1, 0, n_m, 0), \quad \{e_x^\mu\} = (0, 1, 0, 0), \quad A_0 = F_0/k_0, \quad (6)$$

where $\omega_0 = 2\pi\nu_0$ is the circular frequency and $F_0$ denotes the amplitude of the electric field strength. This is an $x$-polarized wave which propagates in the positive $y$-direction in the medium, i.e. the explicit form of the argument of the cosine is $\xi = k \cdot x = \omega_0(t - n_m y/c)$. In this case the matrix part $\sigma \cdot F = 2i(\gamma \cdot k)[\gamma \cdot A'(\xi)]$ in (5c) is proportional with $(\gamma \cdot k)(\gamma \cdot e_x) = -k_0(1 + n_m \alpha_y)\alpha_x$, which is a complex expression in the standard representation (see footnote 1). In the Majorana representation[2] however, we have a *real matrix* for this combination, which considerably simplifies the formulae,

$$(\gamma \cdot k)(\gamma \cdot e_x)/k_0 = +(1 + n_m \beta)\alpha_x = \begin{pmatrix} 0 & (1+n_m)\sigma_x \\ (1-n_m)\sigma_x & 0 \end{pmatrix} = \begin{pmatrix} 0 & 0 & 0 & 1+n_m \\ 0 & 0 & 1+n_m & 0 \\ 0 & 1-n_m & 0 & 0 \\ 1-n_m & 0 & 0 & 0 \end{pmatrix}. \quad (7a)$$

The eigenvalue equation $(1 + n_m \beta)\alpha_x u = \lambda u$ of the above matrix (7a), has the four (normalized) solutions

$$u_1 = \frac{1}{\sqrt{2}}\begin{Bmatrix} +\sqrt{1+n_m} \\ 0 \\ 0 \\ \sqrt{1-n_m} \end{Bmatrix}, \quad u_2 = \frac{1}{\sqrt{2}}\begin{Bmatrix} 0 \\ +\sqrt{1+n_m} \\ \sqrt{1-n_m} \\ 0 \end{Bmatrix}, \quad u_3 = \frac{1}{\sqrt{2}}\begin{Bmatrix} -\sqrt{1+n_m} \\ 0 \\ 0 \\ \sqrt{1-n_m} \end{Bmatrix}, \quad u_4 = \frac{1}{\sqrt{2}}\begin{Bmatrix} 0 \\ -\sqrt{1+n_m} \\ \sqrt{1-n_m} \\ 0 \end{Bmatrix}, \quad (7b)$$

which belong to two real, twofold degenerate eigenvalues $\lambda = \pm\sqrt{1-n_m^2}$,

$$\lambda_1 = \lambda_2 = +\sqrt{1-n_m^2}, \quad \lambda_3 = \lambda_4 = -\sqrt{1-n_m^2} \quad (n_m < 1). \quad (7c)$$

The eigenvectors $u_1$ and $u_2$ (similarly $u_3$ and $u_4$) associated to the common eigenvalue are orthogonal to each other, moreover, $u_1$ is also orthogonal to $u_4$, and $u_2$ is orthogonal to $u_3$. However, $u_1$ and $u_3$ are not orthogonal, and $u_2$ and $u_4$ are not, either. Anyway, this system is complete, and can be orthogonalized by the standard procedure. In the pure electric field case ($n_m \to 0$) the system (7b) itself already forms an orthonormalized basis.

---

[2] The Dirac matrices in the Majorana representation are obtained from the standard representation by using the unitary (and also self-adjoint) transformation matrix $U_M = (\beta + \alpha_y)/2^{1/2}$, for which $U_M^{-1} = U_M^+ = U_M$. In this representation $\alpha'_{x,z} \equiv U_M \alpha_{x,z} U_M^{-1} = -\alpha_{x,z}$, $\alpha'_y \equiv U_M \alpha_y U_M^{-1} = \beta$, $\beta' \equiv U_M \beta U_M^{-1} = \alpha_y$.

It is clear that if $\Psi_p^{(\pm)}$ in (5c) is proportional to one of the eigenvectors $u_{1,2}$ or $u_{3,4}$, then the matrix containing the spin interaction term in (6) can be replaced by a scalar factor, corresponding to the eigenvalues $\lambda_{1,2} = +\sqrt{1-n_m^2}$ or $\lambda_{3,4} = -\sqrt{1-n_m^2}$, respectively. More generally, a solution of (5c) can always be written as an expansion in terms of the eigenvectors $u_{1,2}$ and $u_{3,4}$,

$$\Psi_p^{(\pm)} = \sum_{s=1}^{4} \Psi_{ps}^{(\pm)} u_s, \qquad (7d)$$

and then the scalar coefficient functions $\Psi_{ps}^{(\pm)}$ will satisfy the following four second-order scalar differential equations,

$$\frac{d^2 \Psi_{ps}^{(\pm)}}{dz^2} + (\theta_0 + 2\theta_1 \cos 2z + 2\theta_2 \cos 4z \pm 2i\vartheta_1 \sin 2z)\Psi_{ps}^{(\pm)} = 0, \quad (s = 1, 2, 3, 4) \qquad (8)$$

$$\theta_0 \equiv \frac{4}{-k^2}[-(k \cdot p)^2/k^2 + p^2 - \kappa^2 - \varepsilon^2 A_0^2/2], \quad 2\theta_1 \equiv \frac{4}{-k^2}[\pm 2p_x \varepsilon A_0], \quad z = \xi/2, \qquad (8a)$$

$$2\theta_2 \equiv \frac{4}{-k^2}[-\varepsilon^2 A_0^2/2], \quad \theta_2^{1/2} \equiv \frac{|\varepsilon| A_0}{k_p}, \quad k_p \equiv \sqrt{k^2} \equiv \frac{\omega_p}{c}, \qquad (8b)$$

$$2\vartheta_1 \equiv \frac{4}{-k^2}(\varepsilon A_0 k_0)\sqrt{1-n_m^2} = -\frac{\varepsilon}{|\varepsilon|} 4\theta_2^{1/2}. \qquad (8c)$$

In (8a) we have defined the independent variable $z = \xi/2$ (as is usual in the theory of differential equations with periodic coefficients), and followed the standard notations in introducing the parameters $\theta_{0,1,2}$. We should keep in mind that in (8) the ambient sign $\pm$ in front of the coefficient $\vartheta_1$, refers to the different eigenvalues given by (7c). We also note that without this term (which stems from the EM-field-spin interaction) each equation for $\Psi_{ps}^{(\pm)}$ in (8) would be a real Whittaker–Hill type equation, corresponding to the wave equation of a Klein–Gordon particle (Varró 2013). This mathematical background have been used by Narozhny and Nikishov (1974, see equation (24) in their paper). We also note that the differential equations in our equation (8) for spinor particles have long been used by Nikishov (1970) and by Narozhny and Nikishov (1974, see their equation (39)), where they considered the eigenvalue problem, by using Whittaker's method, and presented analytic perturbative results.

As has already been mentioned above, the standard procedure to solve equations like (8) in terms of a trigonometric series would lead to five-term recurrence relations between the coefficients, in contrast to the three-term expressions, encountered in the case of the Mathieu equation. In looking for finite-term solutions, we overcome this difficulty by using the transformation, originally due to Ince (see references in Arscott 1964). This is the basic new element in our present approach. In the spirit of this method, in order to solve the complex equation in (8), we proceed by introducing the following Ansatz for $\Psi_{ps}^{(\pm)}$

$$\Psi_{ps}^{(\pm)} = f \exp(-\theta_2^{1/2} \cos 2z), \tag{9a}$$

$$\frac{d^2 f}{dz^2} + 4\theta_2^{1/2} \sin 2z \frac{df}{dz} + [\theta_0 + 2\theta_2 + (2\theta_1 + 4\theta_2^{1/2})\cos 2z \pm 2i\vartheta_1 \sin 2z]f = 0, \tag{9b}$$

where the parameters $\theta_{0,1,2}$ have already been defined in (8a-b-c). Henceforth we take a negative test charge (electron), $\varepsilon < 0$ in (8a-b-c), and consider 'positive solutions'. We show the details of the derivation only for the coefficient functions $\Psi_{p1}^{(+)}$ and $\Psi_{p2}^{(+)}$ of the spinor eigenvectors $u_1$ and $u_2$ in (8), which belong to the eigenvalue $+\sqrt{1-n_m^2}$. In this case, from (9b) we obtain

$$\frac{d^2 f}{dz^2} + a \sin 2z \left(\frac{df}{dz} + if\right) + (\eta - qa\cos 2z)f = 0, \tag{10a}$$

$$a \equiv 4\theta_2^{1/2} = 4|\varepsilon|A_0/k_p, \quad q \equiv 2p_x/k_p - 1, \quad 2p_x = (q+1)k_p, \quad k_p \equiv k_0\sqrt{1-n_m^2} \equiv \omega_p/c, \tag{10b}$$

$$\eta \equiv 4[p_\xi^2/k_p^2 + \varepsilon^2 A_0^2 + \kappa^2 - p^2]/k_p^2, \quad p_\xi = (k \cdot p), \quad \eta = 4[\hat{p}^2 + \boldsymbol{p}_\perp^2 + \kappa^2 + \varepsilon^2 A_0^2]/k_p^2. \tag{10c}$$

We would like to emphasize that (10a) is a mathematically new complex equation, which, to our best knowledge, has not been considered so far. In (10b) we have introduced the new parameters $q$ and $k_p \equiv k_0\sqrt{1-n_m^2} \equiv \omega_p/c$, where the subscript 'p' refers to the word 'plasma', though at the present stage we do not need to specify the nature of the medium (in section 4 we shall deal with this interpretation). In deriving the third equation of (10c) we have used the relations (valid for an arbitrary four vector $p_\mu$)

$$p_0^2 - p_y^2 = p_\xi^2/k_p^2 - \hat{p}^2, \quad p^2 = p_0^2 - p_y^2 - \boldsymbol{p}_\perp^2 = p_\xi^2/k_p^2 - \hat{p}^2 - \boldsymbol{p}_\perp^2, \tag{10d}$$

where $\hat{p}$ and $p_\xi$ have been defined in (5b) and (10c), respectively. We note that the sum of the two terms $\kappa^2$ and $\varepsilon^2 A_0^2$ may be combined to $\kappa_*^2 \equiv \kappa^2 + \varepsilon^2 A_0^2$, which is equivalent to introducing the intensity-dependent mass shift $\Delta m = m_* - m = m\sqrt{1+\mu_0^2} - m$, where $\mu_0 = eF_0/mc\omega_0$ is the well-known dimensionless intensity parameter (see e.g. Brown and Kibble 1964). We also note that equation (10a) is unchanged under the simultaneous substitutions $z \to z + \pi/2$ and $a \to -a$, thus we need not consider the case $\varepsilon > 0$ and $\varepsilon < 0$ separately.

## 3. Periodic, finite solutions of the scalar coefficient functions

For the scalar coefficient functions of the bispinor solutions of the second order Dirac equation, introduced in (8), by using the Ansatz (9a), we have derived (10a), which is in fact a complex generalization of Ince's original equation (according to the terminology suggested by Arscott 1964). Our procedure to follow is analogous to that of Ince's method, which we have extended to treat the complex equation (10a), and the associated complex trigonometric polynomials. We note that, though the procedure itself is analogous to that used by Arscott (1964), the complex wave functions obtained by us have quite different properties to that of the real Ince polynomials (Varró 2013).

Let us first expand the solution of equation (10a) as a complex Fourier series,

$$f = \sum_{r=-\infty}^{\infty} D_r \exp(-2irz) . \tag{11a}$$

After substituting into (10a) we receive the recurrence relations for the unknown coefficients,

$$[\eta - (2r)^2]D_r + \{[2(r-1)-1]-q\}(a/2)D_{r-1} - \{[2(r+1)-1]+q\}(a/2)D_{r+1} = 0 . \tag{11b}$$

By shifting the index $r \to r+1$, and multiplying the resulting equation by $(-1)$, we have

$$[½(q+1)-r]aD_r + [4(r+1)^2 - \eta]D_{r+1} + [½(q-1)+r+2]aD_{r+2} = 0 . \tag{11c}$$

Now, if we assume that $½(q+1) = n \geq 1$ is a positive integer, then $½(q-1) = n-1$ is also an integer in the third term in (11c). Moreover, if we assume that $\eta$ has been chosen so that $D_{n+1} = 0$, then the factor in the first term on the left hand side of (11c), with $r = n$, is also zero, regardless of the value of $D_{r=n}$. Accordingly, we have $D_{n+2} = 0$, too. Thus, by using the recurrence relation for $r = n+1, n+2,...$, successively we derive $D_{n+3} = D_{n+4} = ... = 0$, which means that the series terminates with the term $D_n \exp(-2inz)$. By 'going backward', down to the negative values of $r$, and using the same reasoning with $D_{-n} = 0$, one can show that the $D_r$ for all $r < -n+1$ are also zero. Thus, if $½(q+1) = n \geq 1$ then the (potentially infinite) series contains only a finite number ($2n$) of terms with coefficients $D_{-n+1}, D_{-n+2},..., D_0, D_1,..., D_n$. According to the definition of $q$ in (10b), we have found that if the transverse momentum component $p_x$ is an integer multiple of the 'plasma momentum' $k_p$ (i.e. $p_x = ½(q+1)k_p = nk_p$), then there are finite-term, polynomial solutions of the second order Dirac equation. In this case the finite set of recurrence relations are

$$[4((-n)+1)^2 - \eta]D_{-n+1} + [(n-1)+(-n)+2]aD_{-n+2} = 0 , \tag{12a}$$

$$[(n)-r]aD_r + [4(r+1)^2 - \eta]D_{r+1} + [(n-1)+r+2]aD_{r+2} = 0 \quad (-n+2 \leq r \leq n-2) , \tag{12b}$$

$$[(n)-(n-1)]aD_{n-1} + [4(+n)^2 - \eta]D_n = 0 . \tag{12c}$$

The system of equations (12a-b-c) represents the algebraic eigenvalue problem of a tri-diagonal $(2n) \times (2n)$ matrix,

$$\begin{bmatrix} 4(-n+1)^2 & (+1)a & 0 & 0 & 0 \\ (2n-1)a & 4(-n+2)^2 & \ldots & 0 & 0 \\ 0 & (2n-2)a & \ldots & (2n-2)a & 0 \\ 0 & 0 & \ldots & 4(n-1)^2 & (2n-1)a \\ 0 & 0 & 0 & (+1)a & 4n^2 \end{bmatrix} \cdot \begin{bmatrix} D_{-n+1} \\ D_{-n+2} \\ \vdots \\ D_{n-1} \\ D_n \end{bmatrix} = \eta \cdot \begin{bmatrix} D_{-n+1} \\ D_{-n+2} \\ \vdots \\ D_{n-1} \\ D_n \end{bmatrix}. \qquad (13a)$$

The eigenvalues $\eta$ determine through equations (6), (7) and (10c) the possible values of the momentum-like constant of motion $\hat{p}$. By using the notations introduced by Arscott (1964), the tri-diagonal matrix in (13a) is written as

$$\left\langle \begin{array}{ccccc} & (+1)a & \ldots & (2n-2)a & (2n-1)a \\ 4(-n+1)^2 & 4(-n+2)^2 & \ldots & 4(n-1)^2 & 4n^2 \\ (2n-1)a & (2n-2)a & \ldots & (+1)a & \end{array} \right\rangle = \boldsymbol{M}_{2n}(a). \qquad (13b)$$

The eigenvalue equation (13a) and the associated characteristic polynomial then takes the form

$$\boldsymbol{M}_{2n}(a) \cdot \boldsymbol{D}_{2n} = \eta \cdot \boldsymbol{D}_{2n}, \quad \boldsymbol{D}_{2n}^T = [D_{-n+1}, D_{-n+2}, \ldots, D_{n-1}, D_n], \quad \mathrm{X}_{2n}(\eta, a) = \det[\boldsymbol{M}_{2n}(a) - \eta \cdot \boldsymbol{I}_{2n}], \qquad (13c)$$

where the superscript $T$ denotes transpose, and $\boldsymbol{I}_{2n}$ is the $(2n) \times (2n)$ unit matrix. The $(2n)$ eigenvalues are given by the zeros of the characteristic polynomial $\mathrm{X}_{2n}(\eta,a) = \prod_{k=1}^{2n}(\eta - \eta_n^{(k)})$. For such tri-diagonal matrices like our $\boldsymbol{M}_{2n}(a)$, the eigenvalues are all real and different, they are all simple roots of the equation $\mathrm{X}_{2n}(\eta,a) = 0$ (this follows from Lemma 1.(a) stated in subsection 1.8 of Arscott 1964). There are $(2n)$ real linearly independent vectors $\{D_r^{(k)}\}$, associated to the eigenvalues $\eta_n^{(k)}$ $(1 \leq k \leq 2n)$. The general solution (11a) of (10a) reduces to a complex trigonometric polynomial, whose coefficients are the components of the eigenvectors, which satisfy the eigenvalue equation in (13c),

$$f = g_n^k(\xi \mid a, +) = \sum_{r=-n+1}^{n} D_r^{(k)}(a \mid 2n) \exp(-2irz), \quad z = \xi/2, \quad (n = 1, 2, \ldots), \quad (1 \leq k \leq 2n). \qquad (14)$$

These solution may be called 'even solution' of (11a), where, according to (11b), $a \equiv 4|\varepsilon|A_0/k_p$ and $p_x = \frac{1}{2}(q+1)k_p = nk_p$, with $k_p \equiv \sqrt{k^2} = k_0\sqrt{1-n_m^2}$. In $D_r^{(k)}(a \mid 2n)$ the second argument refers to that there are $2n$ sets of the coefficients (eigenvectors), and the superscript labels the $k$ th such set.

By considering still the case $\varepsilon < 0$ and 'positive solutions', and the coefficient functions $\Psi_{p1}^{(+)}$ and $\Psi_{p2}^{(+)}$ of the spinor eigenvectors $u_1$ and $u_2$ in (8), belonging to the eigenvalue $+\sqrt{1-n_m^2}$, we can construct also 'odd solutions' to (10a), by expanding into the following complex trigonometric series,

$$\psi = \sum_{r=-\infty}^{\infty} D_r \exp[-(2r+1)iz]. \tag{15a}$$

We can essentially repeat the steps, which led to the even solutions above. After substitution into (10a) we receive the recurrence relations for the unknown coefficients,

$$[\eta - (2r+1)^2]D_r + \{[2(r-1)+1] - 1 - q\}(a/2)D_{r-1} - \{[2(r+1)+1] - 1 + q\}(a/2)D_{r+1} = 0, \tag{15b}$$

and by shifting the index $r \to r+1$, and multiplying the resulting equation by $(-1)$, we have

$$(\tfrac{1}{2}q - r)aD_r + \{[2(r+1)+1]^2 - \eta\}D_{r+1} + [\tfrac{1}{2}q + (r+2)]aD_{r+2} = 0. \tag{15c}$$

Now, if we assume that $\tfrac{1}{2}q = n \geq 0$ is a non-negative integer, then $p_x = \tfrac{1}{2}(q+1)k_p = k_p(n + \tfrac{1}{2})$, thus in this case $p_x$ contains a 'zero-point momentum' $\tfrac{1}{2}k_p$, too. By assuming that $\eta$ has been choosen so that $D_{n+1} = 0$, the factor in the first term on the left hand side of (15c), with $r = n$, is also zero, regardless of the value of $D_{r=n}$. Accordingly, we have $D_{n+2} = 0$, too, and then $D_{n+3} = D_{n+4} = ... = 0$, thus the series terminates with the term $D_n \exp[+(2n+1)iz]$. The series terminates with the $(-n)$ th term, too. The system of equations now represents the algebraic eigenvalue problem of a tri-diagonal $(2n+1) \times (2n+1)$ matrix,

$$\begin{bmatrix} [2(-n)+1]^2 & (+1)a & 0 & 0 & 0 \\ (2n)a & [2(-n+1)+1]^2 & \ldots & 0 & 0 \\ 0 & (2n-1)a & \ldots & (2n-1)a & 0 \\ 0 & 0 & \ldots & [2(n-1)+1]^2 & (2n)a \\ 0 & 0 & 0 & (+1)a & [2(n)+1]^2 \end{bmatrix} \begin{bmatrix} D_{-n} \\ D_{-n+1} \\ \vdots \\ D_{n-1} \\ D_n \end{bmatrix} = \eta \cdot \begin{bmatrix} D_{-n} \\ D_{-n+1} \\ \vdots \\ D_{n-1} \\ D_n \end{bmatrix}. \tag{16a}$$

By using the notations for the tri-diagonal matrix in (16a)

$$\left\langle \begin{matrix} & (+1)a & \ldots & (2n-1)a & (2n)a \\ [2(-n)+1]^2 & [2(-n+1)+1]^2 & \ldots & [2(n-1)+1]^2 & [2(n)+1]^2 \\ 2na & (2n-1)a & \ldots & (+1)a & \end{matrix} \right\rangle = N_{2n+1}(a), \tag{16b}$$

the eigenvalue equation (17a) and the associated charcteristic polynomial then takes the form

$$N_{2n+1}(a) \cdot \boldsymbol{D}_{2n+1} = \eta \cdot \boldsymbol{D}_{2n+1}, \quad \boldsymbol{D}_{2n+1}^T = [D_{-n}, D_{-n+1}, \ldots, D_{n-1}, D_n],$$

$$Y_{2n+1}(\eta, a) = \det[N_{2n+1}(a) - \eta \cdot \boldsymbol{I}_{2n+1}], \tag{16c}$$

where $\boldsymbol{I}_{2n+1}$ denotes the $(2n+1) \times (2n+1)$ unit matrix. The $(2n+1)$ eigenvalues are given by the zeros of the characteristic polynomial $Y_{2n+1}(\eta, a) = \prod_{k=1}^{2n+1}(\eta - \eta_n^{(k)})$. For the tri-diagonal matrix $N_{2n+1}(a)$, the eigenvalues are all real, and they are all different simple roots of the equation $Y_{2n+1}(\eta, a) = 0$. There are $(2n+1)$ real linearly independent vectors, associated to these eigenvalues $\eta_n^{(k)}$ $(1 \leq k \leq 2n+1)$. The general solution (15a) of (10a) reduces to a complex trigonometric polynomial, whose coefficients are the components of the eigenvectors, which satisfy the eigenvalue equation in (16a-b-c),

$$f = h_n^k(\xi \mid \alpha, +) = \sum_{r=-n}^{n} D_r^{(k)}(a \mid 2n+1) \exp[-(2r+1)iz], \quad z = \xi/2, \quad (n = 1, 2, ...), \quad (1 \le k \le 2n+1). \quad (17)$$

We note that it is a very remarkable property of these solutions that they contain *half-integer harmonics* of the incoming radiation, i.e. they are not $2\pi$ – periodic, but only $4\pi$ – periodic. There are $(2n+1)$ such linearly independent vectors $\{D_r^{(k)}\}$, associated to the eigenvalues $\eta_n^{(k)}$ $(1 \le k \le 2n+1)$. According to (9b), the coefficient functions $\Psi_{p3}^{(+)}$ and $\Psi_{p4}^{(+)}$ of the spinor eigenvectors $u_3$ and $u_4$ in (8), belonging to the eigenvalue $-\sqrt{1-n_m^2}$, satisfy an analogous equation to (10a), with a negative sign in front of the complex imaginary unit. It can be shown that the corresponding even and odd solutions are just the complex conjugates of the functions given in (14) and (17), respectively. This is because the $D'$ coefficients in their Fourier expansion satisfy exactly the same eigenvalue equations as (13a) or (17a), i.e. $D' = D$ in each cases. In this way, on the basis of (14) and (17), we have

$$f = g_n^k(\xi \mid a, -) = [g_n^k(\xi \mid a, +)]^* = \sum_{r=-n+1}^{n} D_r^{(k)}(a \mid 2n) \exp(+2irz), \quad (n = 1, 2, ...), \quad (1 \le k \le 2n), \quad (18)$$

$$f = h_n^k(\xi \mid a, -) = [h_n^k(\xi \mid a, +)]^* = \sum_{r=-n}^{n} D_r^{(k)}(a \mid 2n+1) \exp[+(2r+1)iz], \quad (n = 1, 2, ...), \quad (1 \le k \le 2n+1). \quad (19)$$

From equations (5a), (8), (9a), (14), (17), (18) and (19) we summarize the finite polynomial solutions of the 'even' and 'odd' scalar functions associated to the second-order Dirac equation (3),

$$\Psi_{p1,2}^{(e)} = \exp[+i(\hat{p}\hat{x} + p_x x + p_z z)] \exp[-(a/4)\cos\xi] g_n^k(\xi \mid a, +), \tag{20a}$$

$$\Psi_{p1,2}^{(o)} = \exp[+i(\hat{p}\hat{x} + p_x x + p_z z)] \exp[-(a/4)\cos\xi] h_n^k(\xi \mid a, +), \tag{20b}$$

$$\Psi_{p3,4}^{(e)} = \exp[+i(\hat{p}\hat{x} + p_x x + p_z z)] \exp[-(a/4)\cos\xi] \{g_n^k(\xi \mid a, +)\}^*, \tag{20c}$$

$$\Psi_{p3,4}^{(o)} = \exp[+i(\hat{p}\hat{x} + p_x x + p_z z)] \exp[-(a/4)\cos\xi] \{h_n^k(\xi \mid a, +)\}^*. \tag{20d}$$

For each $n$ – values (corresponding to a discrete set of transverse momenta $2p_x = (q+1)k_p$), the coefficients of the polynomials $g_n^k$ and $h_n^k$ satisfy the eigenvalue equations (13c) and (16c), respectively. In this way, the eigenfunctions and the eigenvalues $\eta_n^{(k)}$ form a *doubly infinite set* labeled by the integer numbers $n$ and $k$. By taking any linear combinations of $\Psi_{ps}$ of (20a-b-c-d), as is shown in (7d), and applying on them the matrix differential operator $(\gamma \cdot \Pi + \kappa)$, we receive an exact solution of the original Dirac equation (1). At present there is no need to show the details of this straightforward calculations.

We note that a new set of exact solutions of the corresponding Klein-Gordon equation found recently by us (Varró 2013) have a similar structure as $\Psi_{ps}$, namely

$$\Phi_p = \exp[+i(\hat{p}\hat{x} + p_x x + p_z z)]\exp[-(a/4)\cos\xi]\varphi_n^k(\xi|a), \tag{21}$$

where the $\varphi_n^k(\xi|a)$ satisfy the Whittaker-Hill equation (which one obtains from (10a) by leaving out the term with the imaginary factor in front of the sine). The functions $\varphi_n^k(\xi|a)$ are real trigonometric polynomials, which coincide with the so-called Ince polynomials (this terminology has been suggested by Arscott 1964). The interrelations between the Ince polynomials and our new complex polynomials (14), (17), (18) and (19) have not been explored, yet.

## 4. Basic properties and some numerical illustrations of the new exact solutions

The investigation of various possible physical applications of the exact solutions (20a-b-c-d) is out of the scope of the present paper. At the end of the present section we shall only give few numerical illustrations. Before doing so, we summarize the basic mathematical properties of these solutions.

On the basis of the original differential equation (8) it can be shown that the functions in (14), (17), (18) and (19) satisfy the following orthogonality relations

$$\int_{-\pi}^{\pi} e^{-(a/2)\cos\xi}(g_n^k)^* g_n^l d\xi = 0, \quad \int_{-\pi}^{\pi} e^{-(a/2)\cos\xi}(h_n^k)^* h_n^l d\xi = 0 \quad (l \neq k). \tag{22a}$$

We normalize these solutions as follows

$$\int_{-\pi}^{\pi} |g_n^k|^2 d\xi = 2\pi; \quad \sum_{r=-n+1}^{n}[D_r^{(k)}(a|2n)]^2 = 1, \quad \int_{-\pi}^{\pi} |h_n^k|^2 d\xi = 2\pi; \quad \sum_{r=-n}^{n}[D_r^{(k)}(a|2n+1)]^2 = 1. \tag{22b}$$

Each of the solutions (20a-b-c-d) contain the exponential factor $\exp[-(a/4)\cos\xi]$ which can be expanded into an infinite Fourier series (Gradshteyn and Ryzhik 1980 or Abramowitz and Stegun 1965),

$$\exp[-(a/4)\cos\xi] = \sum_{l=-\infty}^{\infty} I_l(a/2)\exp[il(\xi-\pi)] = I_0(a/2) + 2\sum_{l=1}^{\infty} I_l(a/2)\cos[l(\xi-\pi)], \tag{23a}$$

where $I_l(z)$ denote the modified Bessel functions of first kind of order $l$. This means that, though the polynomials $g_n^k$ and $h_n^k$ are finite-term expressions, according to (23a), the wave functions in (21a-d) contain all the higher-harmonics of the fundamental frequency. If $a >> 1$, then this function is peaked at the points $\xi = \pi + 2k\pi$ with an exponentially large contrast, we can approximately represent it as a sequence of „delta-kicks",

$$I_l(a/2) = \frac{e^{a/2}}{\sqrt{\pi a}}[1 + O(1/a)], \quad \exp[-(a/4)\cos\xi] \approx \frac{e^{a/2}}{\sqrt{\pi a}} 2\pi \sum_{l=-\infty}^{\infty} \delta[\xi - (2l+1)\pi] \quad (a >> 1). \tag{23b}$$

The differential equation (10a) has led to a two-parameter eigenvalue problem, where we have considered $a = 4|\varepsilon|A_0/k_p$ as a fundamental parameter, and $2p_x = (q+1)k_p$ and $\eta$ were disposable parameters. We have seen that the condition $q+1 = 2n$ has led to even solutions, and $q+1 = 2n+1$ to odd solutions. In (10c) we have presented two equivalent forms of the eigenvalues $\eta$, expressed in terms of the momentum parameters which are determined by the characteristic equations $X_{2n}(\eta,a) = 0$ and $Y_{2n+1}(\eta,a) = 0$ in (13c) and (16c), respectively,

$$\hat{p} = \pm(k_p/2)\sqrt{\eta - (q+1)^2 - (2p_z/k_p)^2 - (2\kappa/k_p)^2 - (a/2)^2}, \quad 2p_\xi/k_p^2 = \pm\sqrt{\eta - (a/2)^2}. \tag{24}$$

The second equation of a simpler form in (24) has been obtained by imposing (in addition) the free mass-shell condition ($p^2 = \kappa^2$) for the momentum parameter $p_\mu$, where the definition of $p_\xi$ in (10c) have also been used. Not all the eigenvalues give physically acceptable solutions of the Dirac equation. More precisely, if $\hat{p}$ becomes purely imaginary, then the wave functions necessarily contain the exponentially growing factor $\exp[\pm|\hat{p}|(k_0^2/k_p^2)(y - n_m x_0)]$ in space-like direction $\hat{k}$. This would not be an acceptable solution, *except for* the case when the interaction is limited in a finite space-time region.

The fundamental parameter $a$ can be expressed in terms of various combinations of parameters which characterize the applied monochromatic field (shortly; laser field) and the medium. If we are allowed to take the Drude free electron model of a plasma medium, with a dielectric permittivity $\varepsilon_m(\omega) = 1 - \omega_p^2/\omega^2 = n_m^2(\omega)$, then $\omega_p$ in (10b) means the plasma frequency of an underdense electron gas, i.e. $4\pi n_e e^2/m = \omega_p^2 < \omega^2$, where $n_e$ is the electron density. In fact, in this case the electromagnetic plane wave under discussion has the dispersion relation $\omega(k_p) = \sqrt{\omega_p^2 + (ck_p)^2}$. Then the fundamental parameter $a$ can be written as the work done by the electric force along the plasma wavelength divided by the photon energy. The ratio of the photon density and the electron density also naturally appears,

$$a = 4\varepsilon A_0/k_p = 4\frac{eF_0\lambdabar_p}{\hbar\omega_0} = 4\sqrt{(2mc^2/\hbar\omega_0)(n_{ph}/n_e)} = 2\mu_0(2mc^2/\hbar\omega_p), \tag{25a}$$

$$n_{ph} \equiv \frac{I_0}{c\hbar\omega_0} = 2.08\times10^8 \times (S/E_{ph})[cm^{-3}], \quad \mu_0 \equiv \frac{eF_0}{mc\omega_0} = 1.06\times10^{-9}\times S^{1/2}/E_{ph}, \tag{25b}$$

$$\omega_p = \sqrt{4\pi n_e e^2/m}, \quad k_p = \omega_p/c = 1/\lambdabar_p = 2\pi/\lambda_p. \tag{25c}$$

We have used the definition $k_p = k_0\sqrt{1 - n_m^2}$ of the 'plasma wavenumber' in (10b), and introduced the plasma frequency $\omega_p$. For example, the plasmon energy $\hbar\omega_p = 1eV$ corresponds to the electron density $n_e = 7.242 \times 10^{20} cm^{-3}$ (in which case the plasma wavelength equals to $\lambda_p = 1240 nm$). In the numerical expressions for the photon number density $n_{ph}$ and for the well-known 'dimensionless intensity parameter' (nowadays also called 'scaled vector potential') $\mu_0$, the quantities $S = I_0/(W/cm^2)$ and $E_{ph} = \hbar\omega_0/(eV)$ measure the intensity in Watts per square centimeters and the photon energy in electronVolts, respectively. For a Ti:Sa laser with $\lambda_0 = 795 nm$ we have $\hbar\omega_0 = 1.563 eV > \hbar\omega_p$, which means that this radiation freely propagates with a phase velocity $c/n_m$ in the underdense plasma. In this case $a = 2 \times 10^6 \mu_0$ and $\mu_0 = 6.782 \times 10^{-10} \times S^{1/2}$, showing that for optical fields the parameter $a$ is by 6 orders of magnitude larger than the usual intensity parameter $\mu_0$. Even for a relatively moderate intensity of $100 MW/cm^2$ ($S = 10^8$), we have already $a = 13.56 \gg 1$. In a recent experiment Kiefer *et al* (2013) have used a Ti:Sa laser of intensity $I_0 = 6 \times 10^{20} W/cm^2$, in which case $\mu_0 = 16.61$ is well beyond the relativistic limit ($\mu_0 = 1$), and consequently $a = 3.32 \times 10^7$ is enormously large. In Figure 1 we present an illustration for the eigenvalue spectrum and one of the corresponding wavefunctions in the former case of a moderate intensity.

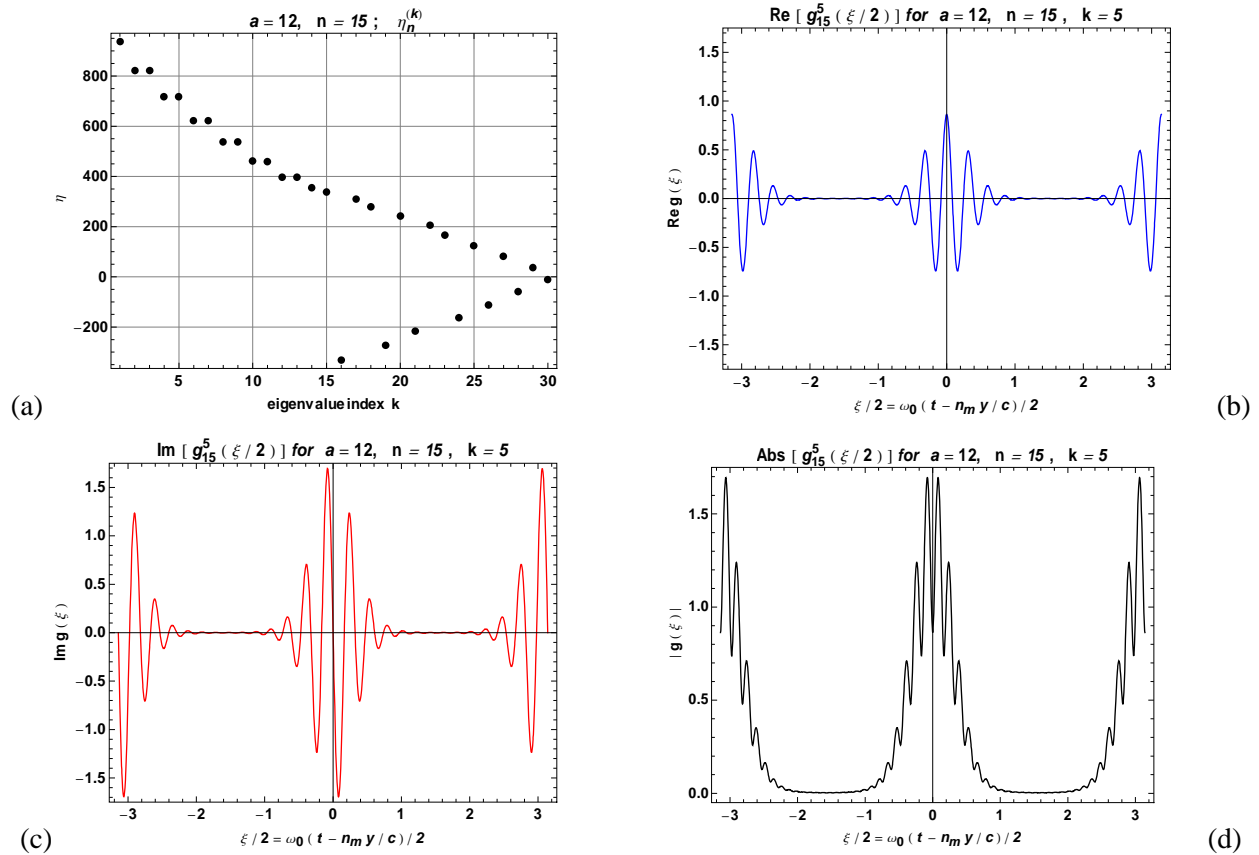

VarroS_Exact-Solution-Dirac_v4x-KCikk, New exact solutions of the Dirac equation of a charged particle interacting with an electromagnetic plane wave in a medium. [ April-May 19. 2013 ]

**Figure 1.** Shows the eigenvalues (a), which are the roots of the characteristic polynomial in (13c), for $a = 12$. The interaction with a Ti:Sa laser field of photon energy $\hbar\omega_0 = 1.563 eV$ and peak intensity $I_0 = 100 MW/cm^2$ has been considered as an example. We have taken $\hbar\omega_p = 1eV$ and $n = 20$, i.e. the transverse momentum of the electron (in original units) is $p_x = 20 \times \hbar k_p$, where $k_p = \omega_p/c$. In the interval $-2\pi \leq \omega_0(t - n_m y/c) \leq 2\pi$, the real part (b), imaginary part (c) and the absolute value (d) of the wave function $g_n^k(\xi | a,+)$ in (14), associated to the fifth eigenvalue $\eta_{15}^{(5)} = 718.092858484742...$ ($k = 5$) are shown. In Figure 1 (a) one sees that the lower index eigenvalues appear in pairs, those with $k =$ (2-3), (4-5), (6-7) and (8-9) cannot be distinguished by eye in the figure. In fact they differ, but only from their 12th decimal digits. That is why we have displayed it with 12 decimal digits, though this accuracy is of no importance here.

According to (24) the parameter $p_\xi$ ( which reduces to the energy $p_\xi/k_0 \to p_0$ in the limit $n_m \to 0$ of pure electric field case) can only be real if $\eta \geq a^2/4$. Among the positive values, this condition still holds for $\eta_{15}^{(29)} > 36$. For these eigenvalues, $p_\xi/k_0 = p_0 - n_m p_y = \pm(k_p^2/2k_0)\sqrt{\eta_n^{(k)} - a^2/4}$ may correspond to a sort of 'gap states', because it can well happen that $-mc^2 < \hbar c p_0 < +mc^2$ in the pure electric case.

In Figure 2 we show the distribution of the harmonic strengths $[D_r^{(k)}(a | 2n)]^2$ in the expansion of the polynomial factors, in case of the interaction with a Ti:Sa laser field of photon energy $\hbar\omega_0 = 1.563 eV$ and peak intensity $I_0 = 100 MW/cm^2$ in a plasma medium of electron density $n_e = 7.242 \times 10^{20} cm^{-3}$ ($\hbar\omega_p = 1eV$). Of course, the complete physical spectrum may in principle be influenced by the exponential prefactor leading to the modified Bessel function series in (23a), but in the present paper we would like to the emphasize on the properties of the by now unknown polynomial factors. According to Figure 2, these spectra are in general qualitatively different from each other.

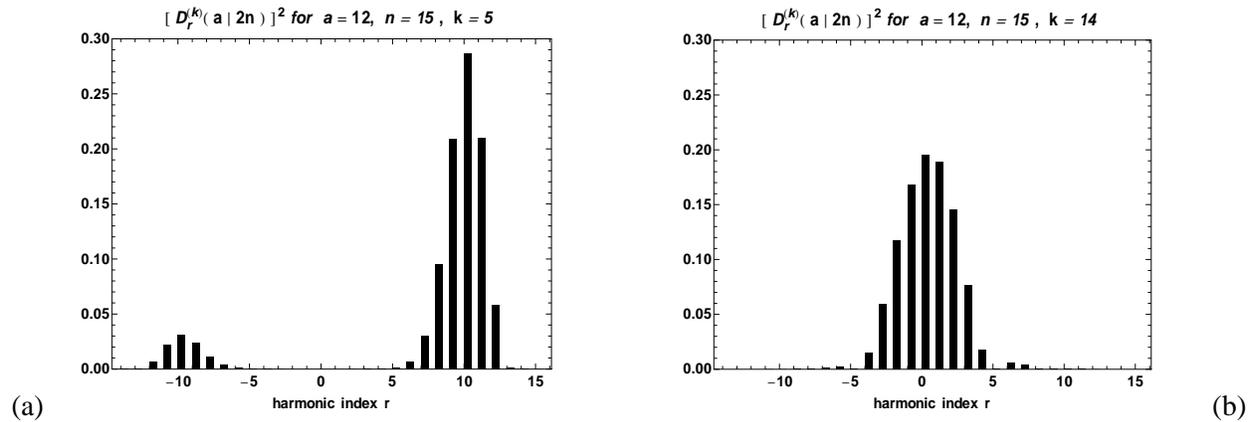

(a)  (b)

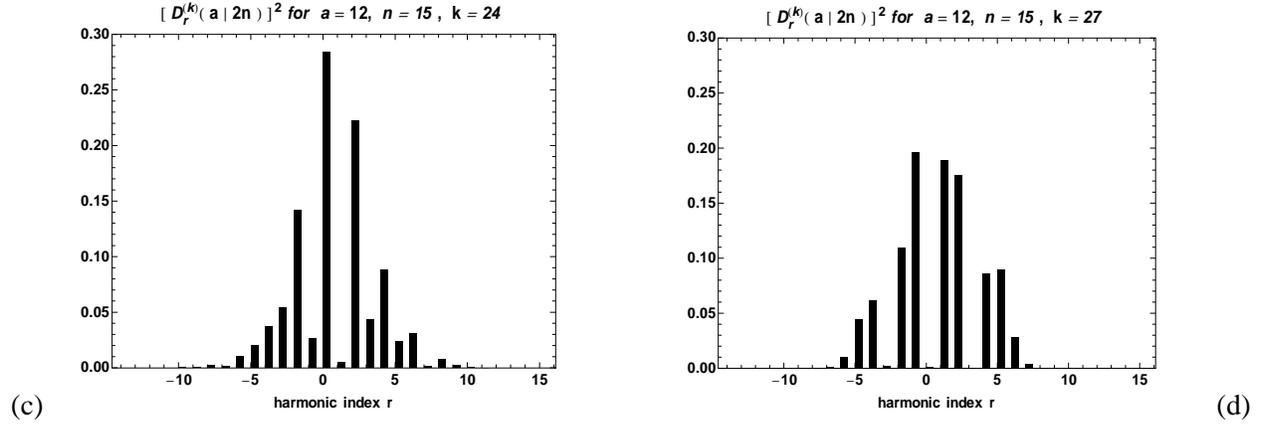

(c)      (d)

**Figure 2.** Shows the strength $[D_r^{(k)}(a|2n)]^2$ of the harmonic coefficients of the polynomial $g_n^k(\xi|a,+)$ for four eigenvalues labelled by the upper index $k = 5, 14, 24, 27$. These eigenvalues are $\eta_{15}^{(5)} = 718.1...$, $\eta_{15}^{(14)} = 355.5...$, $\eta_{15}^{(24)} = -163.1...$, $\eta_{15}^{(27)} = 81.6...$. We are considering the case when $a = 12$ and $n = 15$, and the input parameters are the same as in Figure 1, where the temporal shape of $g_{15}^5(\xi|a,+)$ has been shown.

Finally, we give an overview of the harmonic spectra in Figure 3. The further numerical study of the exact solutions (20a-b-c-d) is out of the scope of the present paper. The discussion of their possible physical applications is also left for future work.

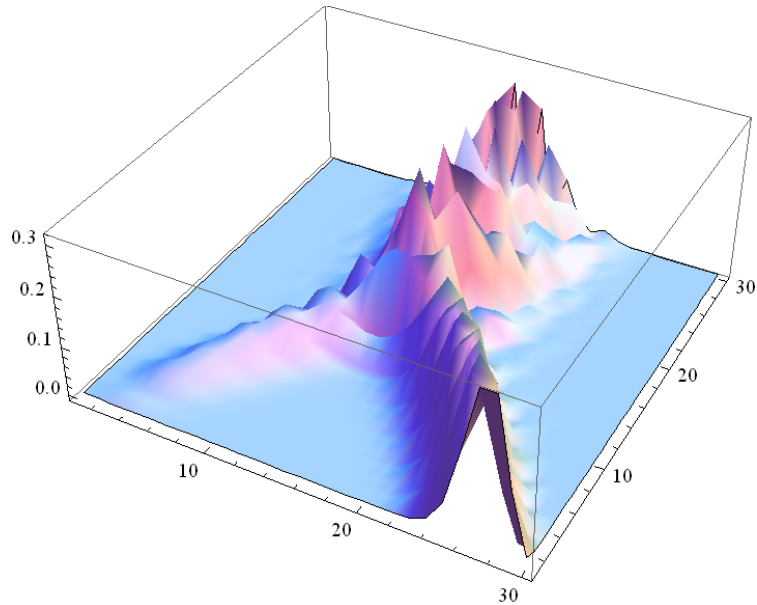

**Figure 3.** Shows an overview of the strength $[D_r^{(k)}(a|2n)]^2$ of the harmonic coefficients on a three-dimensional list plot when $a = 12$ and $n = 15$. This figure summarizes the behaviour of these quantities for different eigenvalues (of indeces $k = 1, 2, ..., 30$, which are drawn on the right axis). The discrete points are connected by a smoothened surface in order to guide the eye. For the lowest index $k = 1$ the distribution is concentrated to positive $r$-values (left axis) in a single peak, the negative index Fourier coefficients are practically zero. For medium $k$-values a double peak structure develops, and for larger values the two peaks merge to an oscillatory distribution. The character of this oscillations has already been illustrated in Figures 2 (c) and (d).

## 5. Summary

We have presented closed form exact solutions of the Dirac equation of a charged particle moving in a monochromatic classical plane electromagnetic field in a medium of index of refraction $n_m < 1$. The solutions found are finite (complex) trigonometric polynomials which form a doubly infinite discrete set, labeled by two integer quantum numbers, which represent a quantized spectrum of the electron's momentum components along the polarization and along the propagation direction of the laser field. As an example for a physical interpretation of these solutions, we have considered the medium an underdense plasma. In this case one of the quantum numbers $n$ characterizes the transverse momentum $p_x = nk_p$ or $p_x = (n+\tfrac{1}{2})k_p$ of the test electron, where $k_p = \omega_p/c$, with $\omega_p$ being the plasma frequency. The other quantum number determines the possible values of the energy parameter, and in certain ranges it is associated to a sort of 'gap states' in the interval $(-mc^2, mc^2)$. The existence of the 'bound states' we have found, may have relevance concerning possible quantum features of mechanisms of laser acceleration of electrons by high-intensity laser fields in an underdense plasma. The fundamental parameter $a$ (see (25a)) which determines the strength of the interaction is the work done on the electron by the electric force of the laser field along a plasma wavelength divided by the photon energy. This $a$ is a quantum parameter, and it is typically by many orders of magnitudes larger than the usual intensity parameter $\mu_0$ (see (25b)).

In secion 2 we have derived the wave equations for the scalar coefficient functions of the Dirac bispinors, and made a brief comparison with the derivation of the usual Volkov states, and the Mathieu-type wave functions in a medium. In section 3 we have derived the basically periodic, finite solutions of the scalar coefficient functions, which are complex trigonometric polynomials. These are associated to the eigenvalue problem of finite special tri-diagonal matrices (13a-b) and (16a-b). We have restricted our analysis only to these special class of solutions, proportional with polynomial expressions (finite complex Fourier sums), which form a subset of all solutions. In section 4 we have presented the basic mathematical properties (like orthogonality) of these solutions, and have given few numerical illustrations for the temporal shape of the wave functions and for the harmonic spectra which are associated to them. We hope to work out possible physical applications based on our above results in the near future.

**Acknowledgments**
This work has been supported by the Hungarian Scientific Research Foundation OTKA, Grant No. K 104260.

VarroS_Exact-Solution-Dirac_v4x-KCikk, New exact solutions of the Dirac equation of a charged particle interacting with an electromagnetic plane wave in a medium. [ April-May 19. 2013 ]